\documentclass[a4paper, 10pt, twocolumn]{article}

\usepackage{etoolbox}	%
\newtoggle{NAT}
\togglefalse{NAT}	%

\usepackage{titling}
\usepackage[backend=biber, 
	style =authoryear-comp, 
	giveninits=true,
	sorting=nyt,
	maxcitenames = 2,
	uniquelist=minyear,
	doi=\iftoggle{NAT}{false}{true},
	isbn,
	url=false,
	sortcites = false,
	date = year]{biblatex}
\newcommand{\apxref}[2][1]{%
\iftoggle{NAT}{#1}{\ref{#2}}%
}

\usepackage[utf8]{inputenc}	%

\usepackage{lmodern}
\usepackage[T1]{fontenc}
\usepackage[babel=false]{microtype}

\usepackage{amsmath}	%
\usepackage{amssymb}	%
\usepackage[affil-it]{authblk}	%
\usepackage[title]{appendix}	%
\usepackage{widetext}
\usepackage{bm}			%
\usepackage{booktabs}	%
\usepackage[font={small}, labelfont={bf}]{caption}	%

\let\counterwithin\relax
\usepackage{chngcntr}	%
\usepackage{color}	%
\usepackage{csquotes}	%
\usepackage{cuted}		%
\usepackage{paralist}	%
\usepackage{graphicx}	%
\usepackage{mathrsfs}	%
\usepackage[group-separator={\,}]{siunitx}	%
\usepackage{stfloats}	%
\usepackage{subfig}		%
\usepackage{tabularx}	%
\usepackage[normalem]{ulem} 		%
\usepackage{units}		%

\usepackage[default]{gfsneohellenic}
\usepackage[LGR,T1]{fontenc} %

\usepackage[hyperfootnotes=false]{hyperref}		%
\usepackage[all]{hypcap}
\usepackage{xcolor}
\definecolor{dark-red}{rgb}{0.45,0.15,0.15}
\definecolor{dark-blue}{rgb}{0.15,0.15,0.4}
\definecolor{medium-blue}{rgb}{0,0,0.5}
\hypersetup{
    colorlinks, 				%
    linkcolor={dark-red},		%
    citecolor={dark-blue},		%
    urlcolor={medium-blue},		%
}

\newcommand*{\fm}[1]{(\ref{#1})}	%

\newcommand*{\bv}[1]{\mathbf{#1}}	%

\newcommand{\K}{K}
\newcommand{\KF}{\K(\lambda)}
\newcommand{\KM}{\K^\mathrm{M}}
\newcommand{\hatKM}{\hat\K^\mathrm{M}}
\newcommand{\F}{\mathscr{F}}

\renewcommand{\P}{\mathscr{P}}

\newcommand{\N}{\mathcal{N}}

\newcommand{\zaSq}{\bv{\zeta}_{a}^2}
\newcommand{\zaVec}{\bv{\zeta}_{a}}
\newcommand{\zapar}{\zeta_{a\parallel}}

\newcommand{\zsp}{\zeta_{sp}}

\newcommand{\zspVec}{\bv{\zeta}_\mathrm{sp}}
\newcommand{\zt}{\zeta_t}

\newcommand{\ztVec}{\bv{\zeta}_\mathrm{t}}
\newcommand{\ztpar}{\zeta_{t\parallel}}
\newcommand{\ztper}{\zeta_{t\perp}}

\newcommand{\thf}{{\bm\theta}}

\newcommand{\power}{\kappa}

\newcommand{\Xt}{{\bv X}_t}
\newcommand{\Wt}{{\bv W}_t}
\renewcommand{\a}{{\bv a}}
\newcommand{\g}{{\bv g}}
\newcommand{\X}{{\bv X}}
\newcommand{\W}{{\bv W}}
\newcommand{\alphaf}{\bm{\alpha}}
\newcommand{\x}{{\bv x}}
\newcommand{\dt}{{\Delta t}}
\newcommand{\Ddt}{{b^2 \dt}}
\newcommand{\alDt}{{\alphaf \dt}}
\newcommand{\dr}{{\Delta{\bv r}}}
\renewcommand{\r}{{{\bv r}}}
\newcommand{\oDr}{\overline{\dr}}

\DeclareMathOperator{\E}{\mathrm{E}}
\DeclareMathOperator{\e}{\mathrm{e}}

\DeclareMathOperator{\pr}{\mathrm{Pr}}
\DeclareMathOperator{\pdf}{\mathit{p}}

\newcolumntype{Y}{>{\raggedright\arraybackslash}X}
\graphicspath{{.}{./figures/}}

\definecolor{burgundy}{rgb}{0.5, 0.0, 0.13}

\title{Statistical Tests for Force Inference in Heterogeneous Environments
}

\newcommand{\affilPasteur}{Decision and Bayesian Computation, USR 3756 (C3BI/DBC) $\&$ Neuroscience department CNRS UMR 3751, Institut Pasteur, CNRS, Paris, France}

\newcommand{\affilMontpelier}{Infectious Disease Research Institute of Montpellier, CNRS UMR 9004, University of Montpellier, Montpellier, France}

\newcommand{\affilCharlesFabry}{Laboratoire Charles Fabry, Institut d'Optique, CNRS UMR8501, Universit\'e Paris Sud, 91127 Palaiseau Cedex, France}

\newcommand{\correspondingauthors}{Correspondence should be addressed to: alexander.serov@pasteur.fr, francois.laurent@pasteur.fr, christian.vestergaard@pasteur.fr, and jean-baptiste.masson@pasteur.fr}

\author[1,*]{Alexander S.~Serov}
\author[1,*]{Fran\c{c}ois Laurent}
\author[2]{Charlotte Floderer}
\author[3]{Karen Perronet}
\author[2]{Cyril Favard}
\author[2]{Delphine Muriaux}
\author[3]{Nathalie Westbrook}
\author[1,*]{Christian L.~Vestergaard}
\author[1,*]{Jean-Baptiste Masson}
\affil[1]{\affilPasteur}
\affil[2]{\affilMontpelier}
\affil[3]{\affilCharlesFabry}
\affil[*]{\correspondingauthors}

\begin{document}

\maketitle

\begin{abstract}

 We devise a method to detect and estimate forces in a heterogeneous environment based on experimentally recorded stochastic trajectories.
 In particular, we focus on systems modeled by the heterogeneous overdamped Langevin equation.
 Here, the observed drift includes a ``spurious'' force term when the diffusivity varies in space.
We show how Bayesian inference can be leveraged to reliably infer forces by taking into account such spurious forces of unknown amplitude as well as experimental sources of error.
 The method is based on marginalizing the force posterior over all possible spurious force contributions. The approach is combined with a Bayes factor statistical test for the presence of forces. The performance of our method is investigated analytically, numerically and tested on experimental data sets. The main results are obtained in a closed form allowing for direct exploration of their properties and fast computation. The method is incorporated into TRamWAy, an open-source software platform for automated analysis of biomolecule trajectories.

 \textbf{Keywords:} overdamped Langevin equation, Bayesian inference, inverse problems, biomolecule dynamics, It\^o-Stratonovich dilemma, random walks
\end{abstract}

\section{Introduction}

Random walks are %
encountered throughout biology and other domains of science, and so is the associated inverse problem of inferring their properties from experimental data.
Random walkers can be considered probes of their environment, and their recorded trajectories thus contain information on the properties of both the walker and its environment.
In the context of biophysics, the random walkers are typically colloidal particles or biomolecules, but, in a general context, they may, for example, represent the motion along an abstract coordinate of a chemical reaction or the fluctuating price of a stock asset.
Transport of biomolecules within cells~\autocite{Wachsmuth2000},
conformational dynamics of proteins and RNA molecules~\autocite{Best2011},
diffusion of proteins on the DNA~\autocite{Vestergaard2018},
dynamics of nanosized objects in the cytosol~\autocite{Etoc2018},
dynamics of receptors in neurons ~\autocite{Dahan2003, Choquet2018, Schneider2015},
complex random walks in mixed biological environments~\autocite{Thapa2018, Grebenkov2018, Javanainen2017, Norregaard2017},
bacteria performing chemotaxis~\autocite{Masson2012, Wong-Ng2016},
immune-cell dynamics~\autocite{Sarris2012,Sarris2015, Fricke2016},
and directionally persistent cell movement~\autocite{Maiuri2015}
are all examples of cases where biologically relevant information can be extracted from recorded stochastic trajectories.

Empirical systems featuring biomolecule random walks are typically characterized by high heterogeneity, so the inverse problem often translates into inferring the properties of the heterogeneous environment from the trajectories of tracer molecules~\autocite{Cocco2008,Best2010, ElBeheiry2015,Masson2009, Hoze2017,  Chang2015, Neuman2008, Lang2004, Li2016, Sungkaworn2017, Remorino2017, Granik2019, Cherstvy2019, Munoz-Gil2020}.
A paradigmatic model for random walks in such systems is the heterogeneous overdamped Langevin equation~(OLE):
\begin{equation}
 d\Xt = \frac {{\bv f}\left(\Xt\right)}{ \gamma\left(\Xt\right)} dt + \sqrt{2D\left({\Xt}\right)}\;\circ d{\Wt} \, ,
 \label{fm:OLE-introduction}
\end{equation}
which describes the continuous-time dynamics underlying a discrete-time recorded random walk~\autocite{Hoze2017}.
Here $\Xt$ is the tracer's position at time $t$, ${\bv f}({\bv X})$ is the force acting on it in the point ${\bv X}$, $D({\bv X})$ is its diffusivity, $\gamma({\bv X})$ is the viscous friction coefficient, and $\Wt$ is a Gaussian zero-mean continuous-time white noise process with uncorrelated increments and unit variance~\autocite{Knight1981}.
Owing to its simplicity, the OLE~\fm{fm:OLE-introduction} is a popular model for biological random walks, providing an effective mesoscopic description of the dynamics~\autocite{Hoze2017}.
As is often the case for models of biological systems, the OLE is empirically postulated rather than derived from the first principles, since this derivation is complex and requires taking into account many factors, such as the heterogeneity of the environment composition, presence of boundaries \autocite{Lancon2002} and hydrodynamic properties \autocite{Crocker1997}, as well as possible noise correlations~\autocite{Franosch2011, Berg-Sorensen2005}.
We refer the interested reader to
(i)~\textcite{Holcman2017, Sancho2011} for an in-depth discussion of the derivation of the OLE and for a microscopic model of crowded environments,
to (ii)~\textcite{vanKampen1988, Jayannavar1995, Lau2007} for some approaches to the derivation of the equation of motion in media featuring diffusivity or temperature gradients,
and to (iii)~\textcite{Kupferman2004, Berg-Sorensen2005, Franosch2011} for a discussion and experimental measurements of the extent to which Brownian noise is truly uncorrelated.

When the diffusivity $D(\x)$ varies in space, Eq.~\fm{fm:OLE-introduction} is only well-defined after a convention for calculating the (stochastic) integral of the noise term has been defined~\autocite[][p.~236]{vanKampen1981}.
Two well-known examples are the It\^o and Stratonovich conventions.
Each convention leads to a different extra ``spurious'' drift term, which is proportional to the diffusivity gradient~\autocite{vanKampen1981, Lau2007, Yang2012}.
This feature of the OLE is known as the It\^o-Stratonovich dilemma~\autocite{vanKampen1992,Lau2007, Sokolov2010, Farago2014, Farago2016, Regev2016}.
For an experimental illustration of the presence of two physically different components see~\autocite{Volpe2010, Brettschneider2011}.

Any stochastic convention can be used in the~OLE to statistically describe a given experimental random walk.
However, they are not equivalent to an external observer attempting to \emph{interpret} the parameters of the random walk.
Indeed, the two components of the drift in the OLE~--- the spurious force and the non-diffusive force~--- have a different physical nature.
The spurious force is proportional to the diffusivity gradient, hence its value changes when the diffusivity, viscosity, or the temperature of the system change in space (time) or across systems.
On the contrary, the non-diffusive component of the drift does not depend on the diffusivity and represents specific or non-specific interactions. 
The spurious force does not contribute to the equilibrium (Boltzmann) distribution.

Spurious forces are due to the interactions of the tracked particle with the surrounding thermal bath, while the non-diffusive forces represent its interactions with other objects or fields~\autocite[see also the discussion of the ``internal noise'' in][p.~234]{vanKampen1992}.
The separation of interactions between these two groups naturally depends (i)~on the scale on which the system is analyzed and (ii)~on which parts of the environment are included into the thermal bath.
In Sect.~\ref{sect:beads-lattice-simulation} below, we show how in the same simulated system, the drift due to non-diffusive forces on the microscopic scale is perceived as a spurious force on the mesoscopic scale, when the contribution of individual interacting partners can no longer be identified.

For practical applications, it is thus important to develop a method allowing to distinguish between diffusive and non-diffusive forces or at least to develop a test allowing to confirm the presence of the non-diffusive forces on a given scale.
The need for such approaches is further emphasized by the inaccessibility of the equilibrium distributions and of the exact boundary conditions at the nanometer scale in numerous biological setups.

Since the seminal work of~\textcite{Bachelier1900}, the inverse problem of drift and diffusivity inference from random walks has been attracting attention~\autocite{Friedrich2011, Kleinhans2012}, especially in financial applications~\autocite{Cootner1970, Parkinson1980}.
In this article, we address a more specific problem of distinguishing between diffusive and non-diffusive forces, since the value of the latter is generally unknown.
Although the spurious force is proportional to the diffusivity gradient $\lambda \nabla D({\bv X})$, it includes an unknown proportionality factor~$\lambda$.
It is known that for physical systems in equilibrium, described by Boltzmann distribution, $\lambda=1$, but its value is not known in general for out-of-equilibrium systems~\autocite{Hanggi1978, Klimontovich1990, Kupferman2004, Lau2007, Volpe2010, Yang2012, Farago2014, Farago2016, Regev2016}.
Each value of $\lambda$ represents specific symmetries of transition probabilities in these systems~\autocite{Klimontovich1990, Klimontovich1994, Sokolov2010}.
Our goal is hence two-fold:
(i)~to develop a statistical test for the presence of non-diffusive forces,
and (ii)~to infer the posterior distribution for the intensity of the non-diffusive forces while taking into account all possible contributions of the ``spurious'' forces as well as experimental localization errors and motion blur.
The method we introduce here is statistically robust to changes in the spurious force contribution in the~OLE due, for example, to changes of the diffusivity or viscosity. 
We validate our approach on numerical trajectories and demonstrate its efficiency on experimental data.

\section{The It\^o-Stratonovich Dilemma for the Inverse Problem}

In this section we give a brief review of the It\^o-Stratonovich dilemma~\autocite{vanKampen1992}.
Numerous discussions of the dilemma have been focused on
choosing the appropriate integral convention for the forward problem of integrating the OLE in a particular system.
In contrast, we here focus on how the dilemma affects the \emph{inverse problem} of inferring the underlying physical parameters of a model from recorded data.
To underline the generality of the problem, we rewrite the OLE~\fm{fm:OLE-introduction} in the form of a general stochastic differential equation~(SDE):
\begin{equation}
 d\Xt = \a(\Xt) dt + b(\Xt) \circ d\Wt,
 \label{fm:OLE}
\end{equation}
where $\a$ and $b$ are differentiable functions of~$\Xt$.
We will refer to $\a$ and $b$ as the drift and diffusivity respectively.

The integral of Eq.~\fm{fm:OLE} is defined as the limit of Riemann sums
\begin{multline}
 \X_{T} = \X_0 + \lim_{N \to \infty} \left[\sum_{i=0}^{N-1} \a(\X_{\xi_i}) (t_{i+1}-t_i) \right.
 \\
 + \left. \sum_{i=0}^{N-1} b(\X_{\xi_i})(\W_{t_{i+1}}-\W_{t_i}) \right],
 \label{fm:RiemannSumsLimit}
\end{multline}
where each point $\xi_i$ is chosen in the interval $[t_i; t_{i+1}]$.
The standard conventions~--- It\^o, Stratonovich-Fisk and H\"anggi-Klimontovich~--- correspond to $\xi_i=t_i$, $\xi_i=(t_i+t_{i+1})/2$ and $\xi_i=t_{i+1}$ respectively~\autocite{Sokolov2010, Lau2007}.
More generally, $\xi_i$ can be set to any point $\xi_i=t_i+\lambda(t_{i+1}-t_i)$ within the~$[t_i; t_{i+1}]$ interval.
This allows one to rewrite Eq.~\fm{fm:OLE} with any convention $\lambda$ in the It\^o form~\autocite{Volpe2016, Kloeden1999}:
\begin{equation}
 d\Xt = \alphaf(\Xt, \lambda)dt + b(\Xt)d\Wt,
 \label{fm:TotalForce}
\end{equation}
where the total drift~$\alpha$ is the sum of $\a$ and the spurious drift $\lambda b(\X)\bv \nabla b(\X)$:
\begin{equation}
 \alphaf(\X, \lambda) \equiv \a(\X)+\lambda b(\X)\bv \nabla b(\X) \enspace.
 \label{fm:alpha}
\end{equation}

From the perspective of the forward problem, Eq.~\fm{fm:alpha} shows that the often arbitrary choice of the value of $\lambda$ influences the value of the drift $\alphaf$ when $\a$ and $b$ are fixed,~--- this is the essence of the It\^o-Stratonovich dilemma~\autocite{vanKampen1992}.
In the context of the inverse problem, one is given fixed values of $\alphaf$ and $b$ estimated from the recorded trajectories, so different choices of~$\lambda$ result in different estimates of the non-diffusive drift~$\a$.
If the chosen $\lambda$ does not agree with its true value in the empirical system, the resulting estimate of~$\a$ becomes biased.

We emphasize that we do not address here the forward problem, i.e. the question of finding the correct $\lambda$ for a given system~\autocite[discussed at length elsewhere, see][]{vanKampen1992, Lau2007, Sokolov2010, Farago2014, Farago2016, Regev2016}.
The correct $\lambda$ values are often inaccessible in real biological systems.
Instead, we aim to solve the inverse problem of whether non-diffusive forces are observed in the system and to infer their values if the appropriate value of $\lambda$ cannot be determined.
It is an inverse problem with an uncertainty in the underlying physical model.
This ambiguity in~$\lambda$ may stem, for example, from the lack of \emph{a priori} knowledge about the out-of-equilibrium fluxes in the system, noise correlations or the particle density distribution.
In all cases, the method developed below allows one to obtain estimates of the non-diffusive forces and to circumvent the It\^o-Stratonovich dilemma by marginalizing over all possible $\lambda$ values.
The estimates are robust to changes in the spurious force contribution in the~OLE.

Above, we have formulated the main question of this paper from a physical point of view as that of inferring non-diffusive forces, when the correct $\lambda$ is unknown.
It is interesting to note that the same question can also be asked from a purely statistical point of view:
\emph{Given the OLE, does there exist a value of~$0 \leqslant \lambda \leqslant 1$ that would allow to describe the given system with zero non-diffusive forces~$(\bv a = 0)$?}
This would allow to describe the same system with fewer parameters ($D$ and $\lambda$ instead of $D$, $\lambda$, $\bv a$), thus minimizing the description length among all the descriptions proposed by the OLE family~\autocite{Balasubramanian2005, Rissanen1986}. %

From this point of view, the Bayes factor developed below is a Bayesian analog  of the difference in the description lengths between the models with~$\bv a \neq 0$ and~$\bv a = 0$ for the given data.
It evaluates how much more efficient the non-diffusive-force description is, as compared to the spurious-force-only description of the same data.
If the spurious-force description is preferred, as a byproduct, one can calculate the value of~$\lambda$ that provides the most efficient description of the data.

\section{The Bayesian Approach}

Our goal is to discriminate between the following two nested hypotheses:
\begin{itemize}
 \item
       $H_0$: the only forces present are spurious forces due to heterogeneous diffusivity (the null hypothesis).
 \item
       $H_1$: there are other, non-diffusive, forces acting on the random walker in addition to the spurious forces.
\end{itemize}
We use the Bayes factor to decide between these hypotheses~\autocite{Kass1995}.

\subsection{The Bayes factor}

According to Bayes' rule~\autocite{Gelman2004}, the posterior probability $\pr(H_i \mid T)$ of a hypothesis~$H_i$ given data~$T$ is
\begin{equation*}
 \pr(H_i \mid T) = \frac{\pdf(T \mid H_i) \pi(H_i)}{\pdf(T)}.
\end{equation*}
Here, $T$ is a trajectory, $T \equiv \{\r_i\}_{i=1}^n$, or a set of trajectories;
$\pdf(T \mid H_i)$ is the marginal likelihood for the data $T$ to be observed under the hypothesis $H_i$; $\pi(H_i)$ is the prior probability of~$H_i$; and $\pdf(T)$ is the probability to observe $T$ under either hypothesis.
For the two competing hypotheses $H_1$ and $H_0$, the ratio of their posterior probabilities reads
\begin{equation*}
 \frac {\pr(H_1 \mid T)} {\pr(H_0 \mid T)} =
 \frac{\pdf(T \mid H_1)}{\pdf(T \mid H_0)}
 \frac{\pi(H_1)}{\pi(H_0)}.
\end{equation*}
The first fraction on the right-hand side is called the \emph{Bayes factor} for $H_1$ over $H_0$~\autocite{Kass1995}:
\begin{equation*}
 \K \equiv \frac {\pdf(T \mid H_1)} {\pdf(T \mid H_0)} \enspace.
\end{equation*}
Each marginal likelihood $\pdf(T \mid H_i)$ is calculated by marginalizing the corresponding conditional likelihood $\pdf(T \mid \thf_i, H_i)$ over all model parameters:
\begin{equation*}
 \pdf(T \mid H_i) = \int d\thf_i \pdf(T \mid \thf_i, H_i) \pi(\thf_i \mid H_i).
\end{equation*}
For $H_0$, the likelihood $\pdf(T \mid \thf_0, H_0)$ thus depends on 3 parameters: $\thf_0 = \{b^2, \g, \lambda\}$, where $b^2$ is the diffusivity and~$\g \equiv \nabla b$ is the diffusivity gradient.
For $H_1$, the likelihood $\pdf(T \mid \thf_1, H_1)$ additionally includes the drift $\a$, so~$\thf_1 = \{b^2, \g, \lambda, \a\}$.
Note that we treat $\g$ as independent from $b^2$, which allows us to obtain the results in the analytical form. 
This assumption is further discussed in Appendix~\apxref[A1]{apx:diffusivity-inference}.

\vspace{12pt}
\subsection{Likelihood}

The likelihood $\pdf(T \mid \thf_i, H_i)$ is obtained as the fundamental solution of the Fokker-Planck equation corresponding to the~OLE~\fm{fm:OLE}.
However, it cannot in general be obtained analytically.
Instead, one can approximate it locally by assuming that $\alphaf$ and $b$ are constant within small spatial domains \autocite{Masson2009,ElBeheiry2015}.
In this case, the likelihood of observing a set of displacements $\{\Delta \bv r\}$ inside a given domain is~\autocite{Masson2009}:
\begin{multline}
 \pdf( \{\Delta \bv r\}\mid \alphaf, b^2) 
\\
 =
 (2 \pi \Ddt)^{-nd/2}
 \exp\left(
 -\frac{n(\oDr - \alDt)^2 + nV}{2 \Ddt}
 \right).
 \label{fm:likelihood}
\end{multline}
Here the mean displacement $\oDr \equiv \sum_{i=1}^n \dr_i/n$ and the biased sample variance $V \equiv \sum_{i=1}^n |\dr_i - \oDr|^2 / n$  are the sufficient statistics of the model~\autocite{Gelman2004}, and $d$ is the number of dimensions.
The equations below are valid for~$d=1$ and~$d=2$, but the framework can also be extended to~$d=3$.

Note that calculations would be similar if one relaxed the approximation of the locally constant values of $\alphaf$ and $b$. 
Computations would be performed numerically but the analytical explorations such as those of Appendix~\apxref[A3]{apx:undetectable-forces} would not be possible. 
Meanwhile, the assumption of bin independence is paramount to the presented method.

\subsection{Priors}

The likelihood~\fm{fm:likelihood} belongs to the exponential family~\autocite{Gelman2004}.
Therefore, a natural choice for the prior is a conjugate prior for the parameters $\a$ and $b^2$.
Among other advantages, conjugate priors provide a closed form of the posterior distribution.
We furthermore assume a factorized form for the prior distributions for $\lambda$ and the diffusivity gradient $\g$:
\begin{equation*}
 \pi(\a, \lambda, \g, b^2 \mid H_1) \approx \pi(a, b^2 \mid \lambda, H_1)\
 \pi(\lambda) \
 \pi(\g) \enspace.
 \label{fm:g-assumption-1}
\end{equation*}
We have no \emph{a priori} information available about the true value of $\lambda$ other than that $0 \leqslant \lambda \leqslant 1$, so we use the flat prior~$\pi(\lambda) \equiv 1$.
The diffusivity gradient prior is approximated by a delta function $\pi(\g) \equiv \delta(\g - \hat\g)$ centered around its maximum \emph{a posteriori} (MAP) value $\hat{\g}$.
Details of $\hat g$ estimation are given in Appendix~\apxref[A1]{apx:diffusivity-inference}.

Under $H_1$, the full conjugate prior is then~(cf.~\fm{fm:likelihood}):
\begin{strip}
\begin{align}
 \pi(\a, \lambda, \g, b^2 \mid H_1) 
 &\equiv
 A_{d}\;
 \left(\Ddt\right)^{-dn_\pi/2}
 \exp\left(
 -\frac{n_\pi ((\a+\lambda\g)\dt - \bm{\mu}_\pi)^2 + n_\pi V_\pi}{2 \Ddt}
 \right)
 \delta(\g - \hat\g) \enspace,
 \label{fm:H1-prior}
 \\
 \pi(\lambda, \g, b^2 \mid H_0)
 &= A_d
 \left(
 \frac{2\pi}{n_\pi \Delta t^2}
 \right)^{d/2}
 \left(
 b^2 \Delta t\right)^\frac{d(1 - n_\pi)}{2} \exp\left(-\frac{n_\pi V_\pi}{2 \Ddt}
 \right)
 \delta(\g - \hat\g).
 \label{fm:H0-prior}
\end{align}
\end{strip}
where %
$A_d \equiv
 (n_\pi/(2\pi))^{d/2}
 (n_\pi V_\pi/2)^{m(d)}
 \Delta t^{d+1} / \Gamma(m(d))$;
$m(d)\equiv d(n_\pi - 1)/2-1$;
$\bm{\mu}_\pi$, $V_\pi$ and $n$ are the parameters of the prior (called hyper-parameters).
The models $H_0$ and $H_1$ are nested models.
In such case, it is common practice to obtain the $H_0$ prior~(Eq.~\fm{fm:H0-prior}) by integrating the $H_1$ prior over~$\a$~\autocite{Kass1995}.

We further set the hyper-parameters to maximally favor the null model.
More specifically, $n_\pi$ acts as an effective number of prior observations.
The least constraining prior is obtained by setting $n_\pi = 4$ for 1D data and $n_\pi = 3$ for 2D, which are the minimal number of observations, for which the prior is proper~(normalized).
Furthermore, we center the prior on zero force by setting $\bm{\mu}_\pi = \lambda\g\dt$.
The remaining hyper-parameter $V_\pi$ defines the prior distribution for the diffusivity.
Sensitivity of the results to~$u\equiv V_\pi/V$ is \mbox{explored in Appendix~\ref{apx:hyper-test}}.

\subsection{Model evidence and the Bayes factor}
\label{sect:MLBF}
The evidence for the $H_1$ and $H_0$ models is the central ingredient in the Bayes factor.
Given the likelihood ${\pdf( \{\Delta \bv r\}\mid \alphaf(\a,\lambda), b^2)}$ (Eq.~\fm{fm:likelihood}) and prior ${\pi(\a, \lambda, \g, b^2 \mid H_1)}$ (Eq.~\fm{fm:H1-prior}), the evidence for~$H_1$ is calculated by marginalizing  ${\pdf( \{\Delta \bv r\}\mid \alphaf(\a,\lambda), b^2)}  {\pi(\a, \lambda, \g, b^2 \mid H_1)}$ over all the parameters $\thf_1=\{a, b^2, g, \lambda\}$.
This gives
\begin{strip}
\begin{align}
 \pdf( \{\Delta \bv r\} \mid H_1) =
 \frac{A_{d} C_d}{(n + n_\pi)^{d/2}}
 \int_0^1 d \lambda \left( n V + n_\pi V_\pi + \frac{n n_\pi}{n + n_\pi} \left(\oDr - \lambda\g\dt \right)^2 \right)^{-\kappa(d)},
 \label{fm:posterior-H1}
\end{align}
\end{strip}
where
$
 C_d=
 2^{\kappa(d)}
 \Gamma(\kappa(d))
 (2\pi)^\frac{d(1-n)}{2}
 \dt^{-d-1}
$ and
$\kappa(d) \equiv d(n+n_\pi -1)/2-1$.

For $H_{0}$, the likelihood $\pdf( \{\Delta \bv r\}\mid \alphaf(\lambda), b^2)$ is given  by Eq.~\fm{fm:likelihood} with $\alpha(\lambda) = \lambda b\nabla b$, and the prior ${\pi(\lambda, \g, b^2 \mid H_0)}$ by Eq.~\fm{fm:H0-prior}. 
Marginalization of ${\pdf( \{\Delta \bv r\}\mid \alphaf(\lambda), b^2)}  {\pi(\lambda, \g, b^2 \mid H_0)}$ over $\thf_0=\{b^2, \g, \lambda\}$ gives
\begin{multline}
 \pdf(\{\Delta \bv r\} \mid H_0) 
 \\
 =
 \frac{A_{d} C_d}{n^{d/2}}
 \int_0^1 d\lambda \left(
 n V + n_\pi V_\pi + n \left(\oDr - \lambda\g\dt \right)^2\right)^{-\kappa(d)}.
 \label{fm:posterior-H0}
\end{multline}

Expressions~\fm{fm:posterior-H1} and~\fm{fm:posterior-H0} let us finally calculate the marginalized Bayes factor $\KM$, which takes into account all possible values for the unknown parameter $\lambda$.
For comparison, we also provide the Bayes factor~$\KF$ for fixed-$\lambda$ inference procedures (It\^o, Stratonovich or H\"anggi), which is calculated in the same manner:
\begin{align}
 \KM
 &=
 \eta^d
 \frac{
  \int_0^1 d \lambda \;
  \left[
   v + \eta^2(\ztVec - \lambda \zspVec)^2
   \right]^{-\power(d)}
 }
 {
  \int_0^1 d \lambda \;
  \left[
   v + (\ztVec - \lambda \zspVec)^2
   \right]^{-\power(d)}
 },
 \\
 \KF
 &=
 \eta^d
 \left[
  \frac{
   v + \eta^2(\ztVec - \lambda \zspVec)^2
  }
  {
   v + (\ztVec - \lambda \zspVec)^2
  }
  \right]^{-\power(d)}.
 \label{fm:bayes-factor-final}
\end{align}
All the integrals appearing in Eqs~(\ref{fm:posterior-H1}--\ref{fm:bayes-factor-final}) are 1D integrals that are numerically evaluated using the trapezoid rule.

The natural parameter combinations appearing in Eq.~\fm{fm:bayes-factor-final} are: %
(i)~$\ztVec \equiv \oDr/\sqrt{V}$, the signal-to-noise ratio for the total force in a single displacement; %
(ii)~$\zspVec \equiv \hat\g\dt / \sqrt{V}$, the signal-to-noise ratio for the spurious force in a single displacement; %
(iii)~$\eta \equiv \sqrt{n_\pi / (n + n_\pi)}$, the relative strength of the prior compared to the observed data; %
(iv)~$v \equiv 1 + n_\pi V_\pi / (n V)$, a weighted ratio of jump variances in the prior and in the data.

\begin{figure*}[htb]
 \includegraphics[width = 7in]{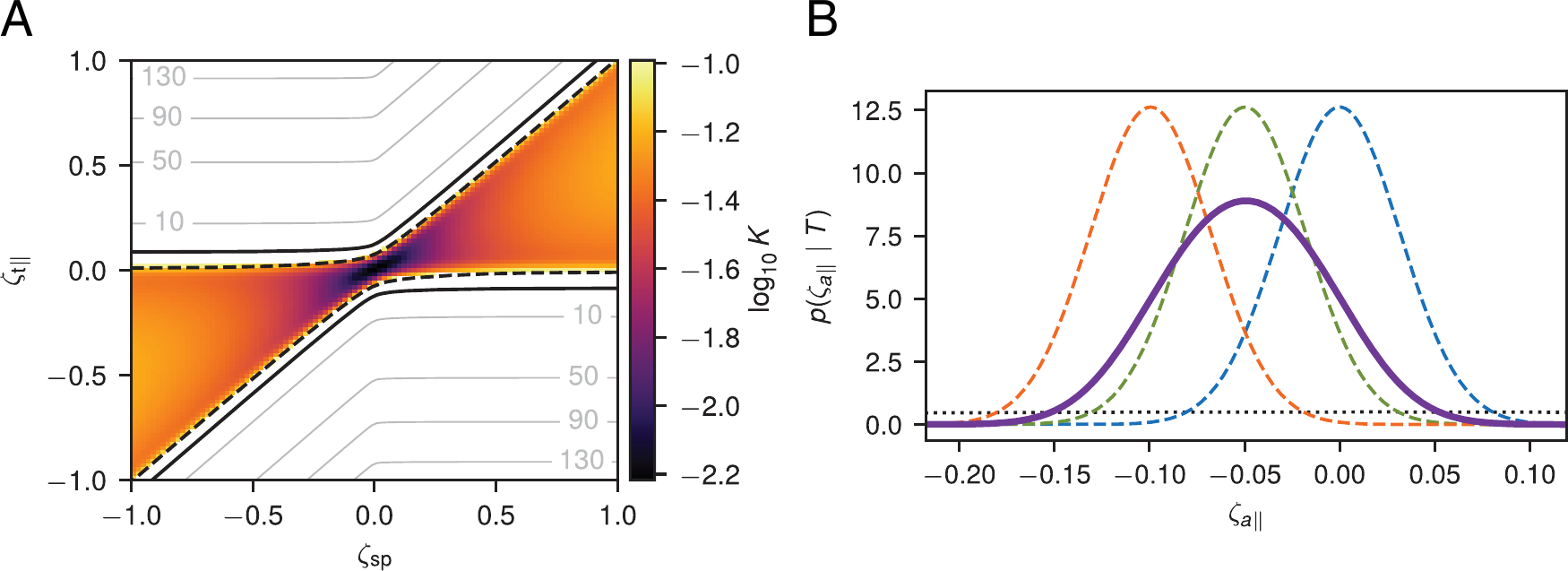}
 \caption{
  \textbf{A}) Marginalized Bayes factor $\KM$ for the presence of forces in a 2D system as a function of the signal-to-noise ratios $\zsp$ and~$\ztpar$~(see text).
  Black lines show the Bayes factor levels $\log_{10}\KM =1$~(solid) and $\log_{10}\KM =-1$~(dashed).
  Gray lines mark levels of Bayes factor for $\log_{10}\KM \geqslant 10$.
  The color map shows Bayes factor values for $\log_{10}\KM \leqslant -1$.
  $\KM$ behavior is qualitatively similar for other parameter values.
  \textbf{B})  Force posteriors for a 2D system with $\zsp = 0.1$ and $\ztVec = 0$.
  Posterior distributions obtained for It\^o (dashed blue line), Stratonovich (dashed green), H\"anggi (dashed orange) and marginalized (solid magenta) approaches are shown alongside their common prior (dotted black).
  In both panels, the number of recorded displacements is $n=500$, the prior hyperparameter~$u = 1.0$, the perpendicular component of the total force~$\ztper = 0$, and the localization error $\sigma_L^2=0$
 }
 \label{fig:figure-1}
\end{figure*}

Figure~\ref{fig:figure-1}A plots the marginalized Bayes factor $\KM$~\fm{fm:bayes-factor-final} as a function of~$\zsp$, and of the component of the total force $\ztpar$ parallel to~$\zspVec$.
The lowest values of $\KM$ are achieved in the region $0 \leqslant \ztpar/\zsp \leqslant 1$.
The value of $\KM$ changes relatively little within this region but grows rapidly  at its boundary.
The absolute minimum of $\KM$ is achieved for~$\zsp=0$ and~$\ztpar=0$ with $\min \KM = \min \KF = \eta^d [(v+\eta^2 \ztper^2) / (v + \ztper^2)]^{-\kappa}$.
A mathematical analysis of Eq.~\fm{fm:bayes-factor-final} is provided in Appendix~\apxref[A3]{apx:undetectable-forces}, where it is shown that non-diffusive forces cannot in principle be detected in certain intervals of $\zspVec$, $\ztVec$ regardless of the number of collected data points.
Appendix~\apxref[A4]{apx:positioning-noise} extends the Bayes factors~\fm{fm:bayes-factor-final} to the experimentally relevant case with localization errors and motion blur.

\subsection{Force posterior}

When $H_1$ is met, we can infer the value of the non-diffusive force by marginalizing the force posterior over all possible values of $\lambda$:
\begin{multline}
 \pdf(\bv \zeta_a \mid T)
 \\
 \propto
 \int_0^1
 d \lambda
 \left[
  v
  + \frac{\eta^2}{1-\eta^2} \zaSq
  + (\ztVec - \zaVec - \lambda \zspVec)^2
  \right]^{-\power(d) - d/2},
 \label{fm:active-force-posterior}
\end{multline}
where a signal-to-noise ratio for the force $\zaVec \equiv \bv a \Delta t / \sqrt{V}$ was introduced.
Figure~\ref{fig:figure-1}B plots an example force posteriors obtained with the marginalized method and with fixed-$\lambda$ inference schemes.
The wider marginalized method posterior takes into account all possible $\lambda$ values.
Appendix~\apxref[A5]{apx:non-symmetric-posterior} demonstrates that in contrast to the fixed-$\lambda$ posteriors, the marginalized posterior is in general non-symmetric.

\subsection{Numerical results}

The performance of the marginalized method was investigated on simulated trajectories.
Random trajectories were simulated in a 2D box with periodic boundary conditions, a uniform total force, and a triangular diffusivity profile along the $x$ axis~(Fig.~\ref{fig:figure-2}A,B).
Other simulation parameters are given in Appendix~\apxref[A6]{apx:numerical-simulation-parameters}.
For each trial, the simulated trajectories were then analyzed using the TRamWAy software platform~\autocite{Tramway2018} and following a procedure similar to the one used in~\textcite{ElBeheiry2015,Remorino2017} and consisting of %
(i)~individual spatial tessellation in each trial; %
(ii)~assignment of recorded displacements to spatial domains; %
(iii)~inference of $\zspVec$ and $\ztVec$ in each domain; %
(iv)~calculation of the Bayes factor in each domain.

\begin{figure*}[htb!]
 \includegraphics[width=7 in]{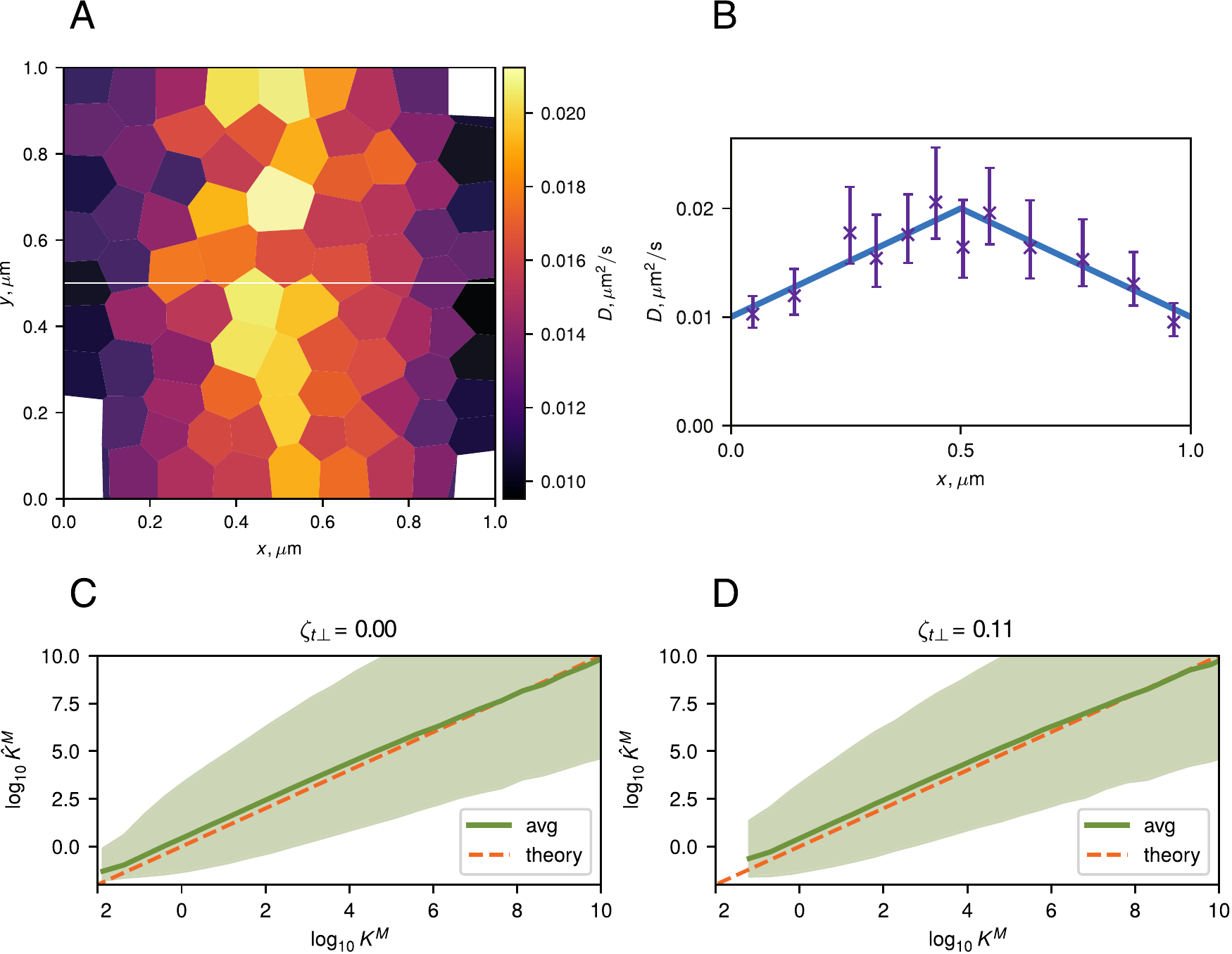}
 \caption{
  \textbf{A}) A diffusivity map inferred using the TRamWAy software platform~\autocite{Tramway2018}.
  The white line $y=\unit[0.5]{\mu m}$ indicates the axis along which the diffusivity profile is plotted in \textbf{B}.
  \textbf{B}) 1D diffusivity profile showing the true diffusivity (solid blue line) and $D$ values inferred from a single simulation (magenta crosses).
  Error bars show \unit[95]{\%} confidence intervals~(CI) calculated from the diffusivity posterior~(see Appendix~\apxref[A1]{apx:diffusivity-inference}).
  \textbf{C,D}) Inferred values of the marginalized Bayes factor $\log_{10}\hatKM$ as a function of its expected value: sliding average (solid green line) over a window of constant width~0.5 on the logarithmic scale and the corresponding \unit[95]{\%} CI (shaded green region).
  The observed dependency is centered around the identity line~($\hatKM = \KM$, dashed orange) indicating that the inferred Bayes factors are approximately unbiased for both the case when the total force is parallel to the spurious force~($\ztper = 0$, \textbf{C}) and when it is not~($\ztper = 0.11$, \textbf{D}).
  The Bayes factors were inferred in individual domains shown in \textbf{A}, with an average of $\sim 123$ individual domains per trial.
  In each trial, the spatial tessellation was performed independently based on the relative particle density.
  The calculations were repeated across 100~trials for each value of~$\ztpar$ out of the analyzed range~(see Appendix~\apxref[A6]{apx:numerical-simulation-parameters})
 }
 \label{fig:figure-2}
\end{figure*}

The marginalized Bayes factor $\hatKM$, inferred in each domain, was then plotted against its expected value $\KM$ to test the accuracy of the method~(Fig.~\ref{fig:figure-2}C,D).
The figure shows good correspondence between the inferred Bayes factor and the expected Bayes factor.
\unit[95]{\%} confidence intervals (CIs) show the extent of the deviation of the results from the true values due to the stochastic nature of the simulated trajectories.

\subsection{Microscopic model of heterogeneous diffusivity}
\label{sect:beads-lattice-simulation}

\begin{figure*}[htb!]
 \includegraphics[width=7 in]{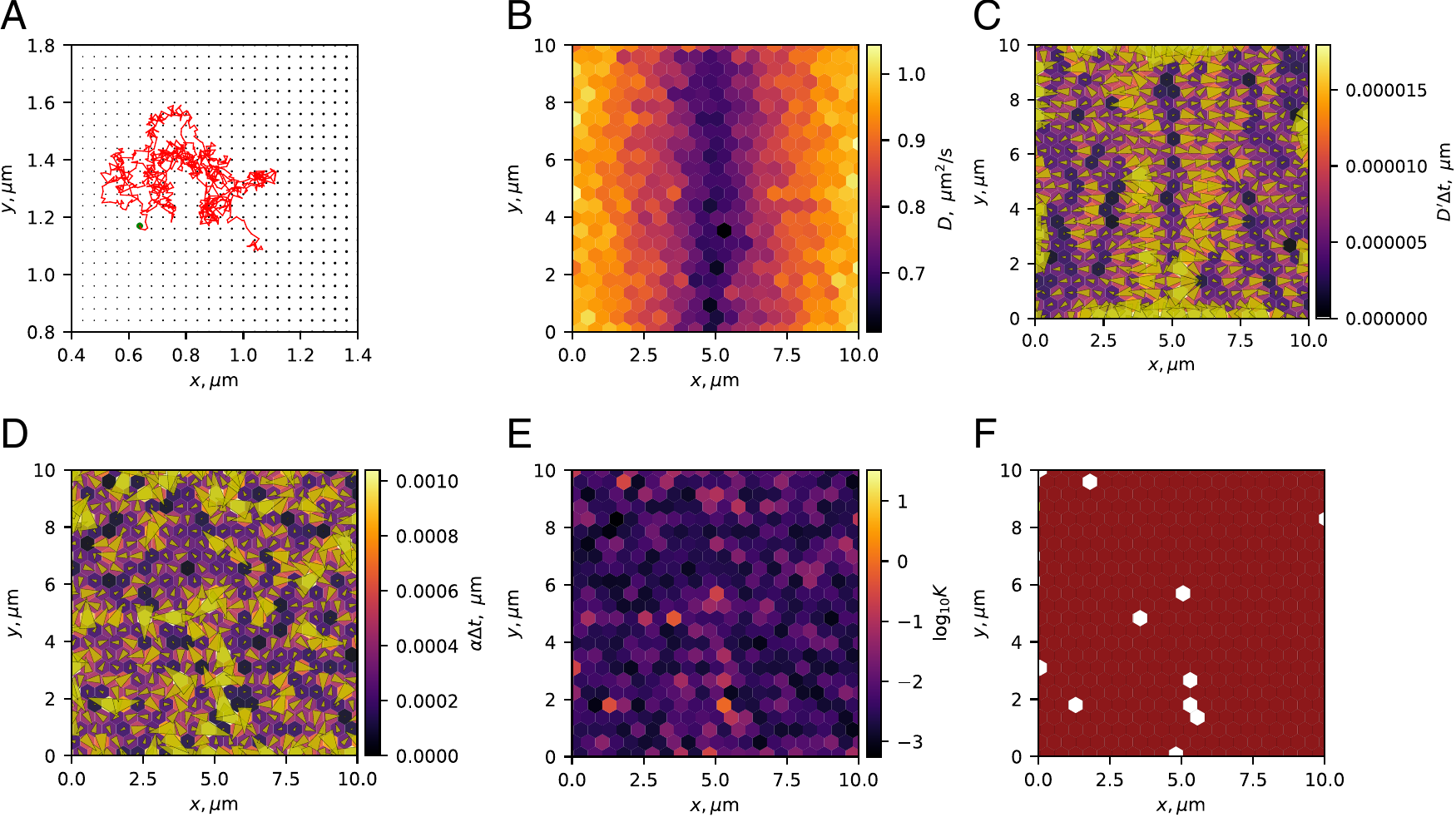}
 \caption{Simulations of mesoscopic changes in particle diffusivity due to microscopic crowding.
  Inference results for particles diffusing in the presence of a lattice of immobile beads of various radius.
  \textbf{A)} A zoom-in on a small $\unit[1 \times 1]{\mu m^2}$ section of the simulated $\unit[10 \times 10]{\mu m^2}$ system.
  The radius of the immobilized beads located in the nodes of a lattice changes with $x$ leading to an effective diffusivity gradient on a larger scale~(cf.~\textbf{B,C}).
  A sample trajectory of 1000~jumps of a single diffusing particle, with a green circle indicating the origin point.
  In total, 1000 independent diffusing particles were simulated.
  \textbf{B)} Inferred diffusivity.
  \textbf{C)} Inferred diffusivity gradient.
  Arrows indicate the direction and the strength of the gradient, also represented by the bin color.
  \textbf{D)} Inferred drift.
  Arrows indicate the direction and the strength of the drift, also represented by the bin color.
  \textbf{E)} Estimated Bayes factor.
  \textbf{F)} Thresholded Bayes factor.
  Color code: green (non-spurious force, $\log_{10} K \geqslant 1$), red (spurious force only, $\log_{10} K \leqslant -1$), white (insufficient evidence, $|\log_{10} K| < 1$).
  Values of $D$, $D'$ and $\alpha$ in the plots (\textbf{B, C, D}) were clipped at high values around the 9th decile to allow for a clearer visualization.
  Simulation details are provided in Appendix~\apxref[A7]{apx:beads-lattice-simulations}
 }
 \label{fig:figure-lattice-of-beads}
\end{figure*}

The next simulation was performed with two goals:
(i)~to illustrate how spurious forces may originate from crowding at the molecular scale,
and (ii)~to illustrate a case, wherein our developed statistical test successfully indicates the absence of non-diffusive forces.
For this purpose, we simulated free diffusion of particles with no microscopic drift within a square region with periodic boundaries and with impenetrable immobile beads evenly spaced on a square lattice~(Fig.~\ref{fig:figure-lattice-of-beads}A), similar to schemes suggested in\autocite{Machta1983, Holcman2011, Chakraborty2019}.
The microscopic diffusivity of the particle was the same throughout the system.
A spatial variation in the radii of the immobile beads created a spatial variation in the effective diffusivity on a much larger ``mesoscopic'' scale, where each analysis bin included $\sim 100$ small beads~(Fig.~\ref{fig:figure-lattice-of-beads}B).
As a result, recordings at the mesoscopic scale exhibit a diffusivity gradient~(Fig.~\ref{fig:figure-lattice-of-beads}C), which contributes to the drift observed on the same scale~(Fig.~\ref{fig:figure-lattice-of-beads}D).
Note that at long time scales, the system is in physical equilibrium, although particles experience a stationary non-zero drift.
Simulation details and parameters are provided in Appendix~\apxref[A7]{apx:beads-lattice-simulations}.

The diffusivity gradient contribution to the drift is the spurious force, its exact value depends on~$\lambda$.
Assuming the value of~$\lambda$ is unknown, one can use the Bayes factor test developed above to estimate the \emph{a posteriori} likelihood of that the observed drift is due to non-spurious forces~(Fig.~\ref{fig:figure-lattice-of-beads}E).
In our simulation, the inferred Bayes factors were small~($\log_{10} K < -1$) in most parts of the region, supporting the claim that only spurious forces were present~(Fig.~\ref{fig:figure-lattice-of-beads}F).
Statistical noise in several bins resulted in weaker evidence, which did not let us draw statistically significant conclusions in those zones.
These results confirm the capacity of the method to detect spurious forces.
Its capacity to detect non-spurious forces will be illustrated in the next section.

This simulation captures one possible microscopic mechanism behind the observation of a diffusivity gradient on the mesoscopic scale in biological systems.
However, note that the homogeneous composition required for a uniform microscopic diffusivity is probably achievable in the biological systems only on the molecular scale~($\unit[10^{-9}]{m}$ and smaller).
On this scale though, it is not clear whether the diffusivity itself is well-defined, since by definition it is the result of the action of millions of individual molecules and Fick's law describes an intrinsically mesoscopic phenomenon.

Other microscopic mechanisms for the diffusivity gradient include
(i)~confinement, wherein it was shown that the diffusivity in a homogeneous system changes with the distance to a wall~\autocite{Brenner1961, Lancon2002, Lau2007, Volpe2010},
(ii)~corralled motion~\autocite{Niehaus2008},
(iii)~hydrodynamic coupling to other objects in the medium~\autocite{Crocker1997, Batchelor1976},
(iv)~temperature gradients~\autocite{Duhr2006, Bringuier2007, Yang2012},
or (v)~intermittent trapping~\autocite{Metzler2014}.

\section{Applications}

The developed method was tested on two experimental systems.
The first one was a well-controlled setup of a bead in the optical tweezers.
The second one was a complex biological process of HIV virion assembly in a T cell~\autocite{Freed2015}, where the OLE is potentially only an approximation to the true biomolecule dynamics~(ignoring inertial effects, colored noise or memory of the previous states).

\subsection{Optical tweezers}
\begin{figure*}[htb!]
 \includegraphics[width=7 in]{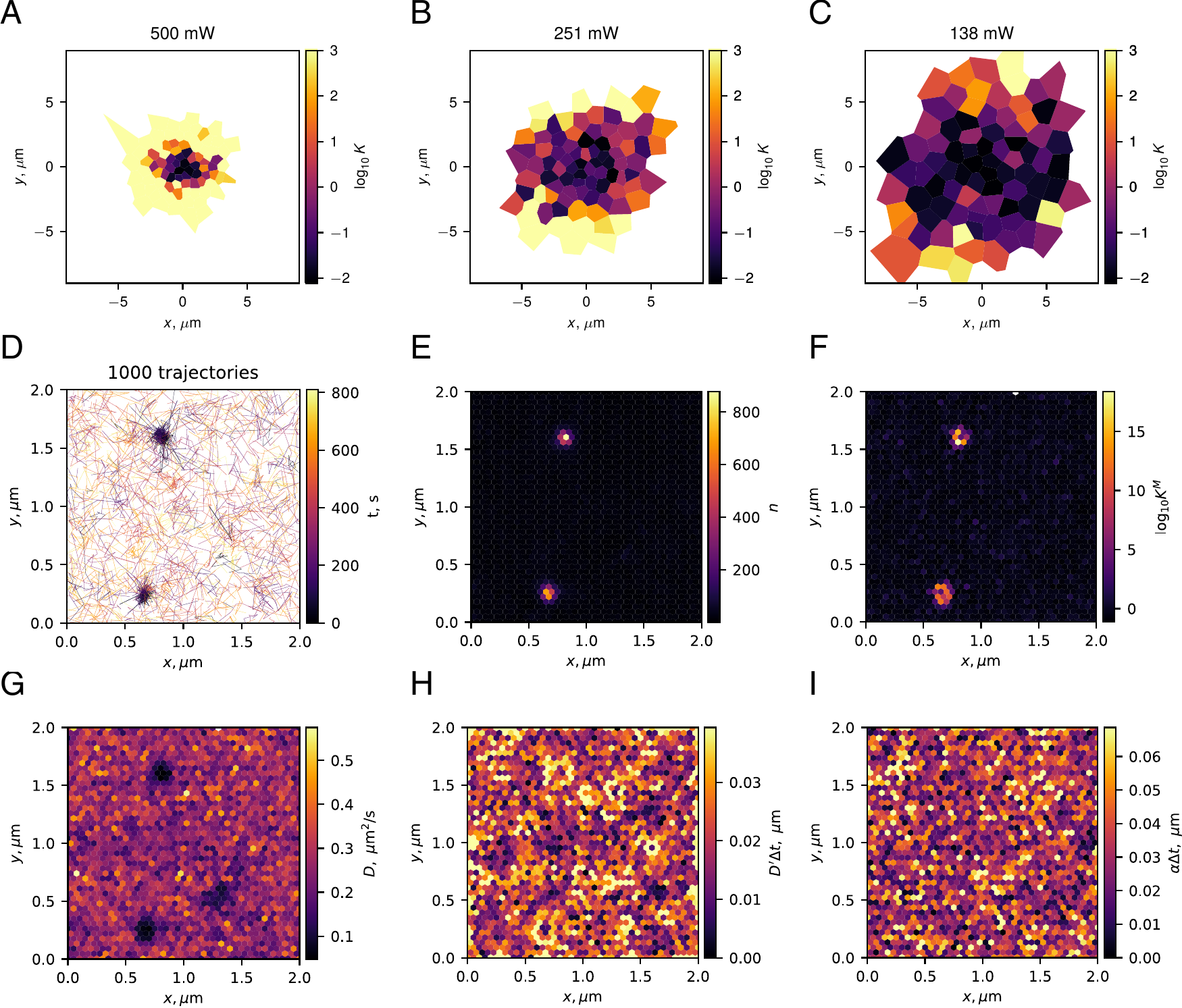}
 \caption{
  \textbf{A--C)} Bayes factors for the presence of non-spurious forces inferred from experimentally-recorded trajectories of a bead trapped in the optical tweezers at 3 different levels of laser power: \textbf{A})~\unit[500]{mW}, \textbf{B})~\unit[251]{mW}, \textbf{C})~\unit[138]{mW}.
  For a better visual representation, all values of $\log_{10} \KM >3$ (very strong evidence for $H_1$) are shown as $\log_{10} \KM = 3$.
  Each domain contained strictly between 390 and~410 recorded displacements.
  \textbf{D--I}) Bayes factors analysis for single-molecule-dynamics of the \emph{Gag} protein during the assembly of HIV VLPs in human T cells.
  \textbf{D--I)} All panels show the same $\unit[2]{\mu m} \times  \unit[2]{\mu m}$ patch of a T cell membrane.
  \textbf{D)}~1000 trajectories randomly chosen among the \num{12825} trajectories of the data set.
  \textbf{E)} Number of displacements $n$ attributed to each domain.
  \textbf{F)} Common logarithm of the marginalized Bayes factor~$\KM$.
  Note the high values of $\log_{10} \KM$ where \emph{Gag} particles cluster.
  \textbf{G)} Inferred diffusivity field.
  \textbf{H)} Absolute value of the inferred diffusivity gradient.
  \textbf{I)} Absolute value of the inferred total force.
  In panels (\textbf{H, I}), large values have been clipped around the 9th decile to increase plot readability
 }
 \label{fig:figure-tweezers-vlp}
\end{figure*}

Optical tweezers combine physical trapping of the bead with local laser heating of the medium, leading to a heterogeneous diffusivity field.
Therefore, the heating effect and the ensuing spurious forces may interfere with the inferred trapping potential.
Figure~\ref{fig:figure-tweezers-vlp}A--C compares the results of Bayes factor calculations for the same system subjected to three different laser powers.
The tessellation procedure was designed to assign the same number of jumps to each domain.
In all 3 cases, the particle is confined and the Bayes factor favors the presence of forces~($\log_{10} \KM > 1$) in a large number of domains, which form a connected region.
With the decrease of the laser power, the confinement at the center of the trap becomes more shallow, so that the statistical test only detects confining forces on the trap border.

\subsection{Assembly of HIV-1 Virus-Like Particles}

The HIV virus-like-particle~(VLP) assembly experiments that provided the data are described in~\textcite{Floderer2018}.
The VLPs derive from the human immunodeficiency virus type~1~(HIV-1), but are immature and deprived of envelope proteins.
One of their main components is the group-specific antigen~(\emph{Gag}) protein.
It is a viral structural protein produced by the virus that anchors and oligomerizes at the plasma membrane of the host T cells, eventually assembling into a VLP~\autocite{Freed2015}.
In the experiments, the HIV-1 \emph{Gag} precursor was genetically modified to contain a photoactivable fluorescent tag mEOS2 protein.
It allowed to record VLP assembly in human CD4 T cells using single-particle tracking photoactivated localization microscopy (sptPALM)~\autocite{Manley2008, Floderer2018}.
Several VLPs can assemble in parallel in the same observation region.

The TRamWAy software platform was used to tessellate the observation region and infer maps of diffusivities~(Fig.~\ref{fig:figure-tweezers-vlp}G) and drift~\autocite[Fig.~\ref{fig:figure-tweezers-vlp}I, see][]{ElBeheiry2015, Tramway2018}.
The Bayes factor map was then computed.
The localization uncertainty was~$\sigma_L= \unit[30]{nm}$, requiring corresponding corrections to the Bayes factor~(Appendix~\apxref[A4]{apx:positioning-noise}).
Inference results and Bayes factors for a $\unit[2]{\mu m} \times \unit[2]{\mu m}$ zone of one T-cell membrane are shown in Fig.~\ref{fig:figure-tweezers-vlp}D-F.

Plots of the trajectories, the density of the recorded points and the diffusivity~(Fig.~\ref{fig:figure-tweezers-vlp}D,E,G) indicate that there are two regions of interest~(ROI) in the data set.
However, the plots of the diffusivity gradient and the drift~(Fig.~\ref{fig:figure-tweezers-vlp}H,I) suggest that the two parameters are of the same scale, hence it is not \emph{a priori} clear, whether the localization of the particles is due to non-diffusive or spurious forces.
Only the calculation of the Bayes factors for these regions allowed us to confirm that it is not solely due to heterogeneities in the diffusivity but rather to non-spurious forces~($\log_{10}(\K)\gg 1$, Fig.~\ref{fig:figure-tweezers-vlp}F).

Some other individual domains in Fig.~\ref{fig:figure-tweezers-vlp}F bear evidence of a force with rather high Bayes factors~($\log_{10}(\K)\geq 1$).
In such a complex system, the high $\K$ values in these individual domains may stem from local membrane activity, failed capsid assembly~\autocite{Floderer2018} or be false detection.
In the rest of the region, the Bayes factor is $|\log_{10} (K)| < 1$ meaning neither of the models is favored at the chosen level of statistical significance.

\begin{figure*}[htb!]
 \includegraphics[width=7 in]{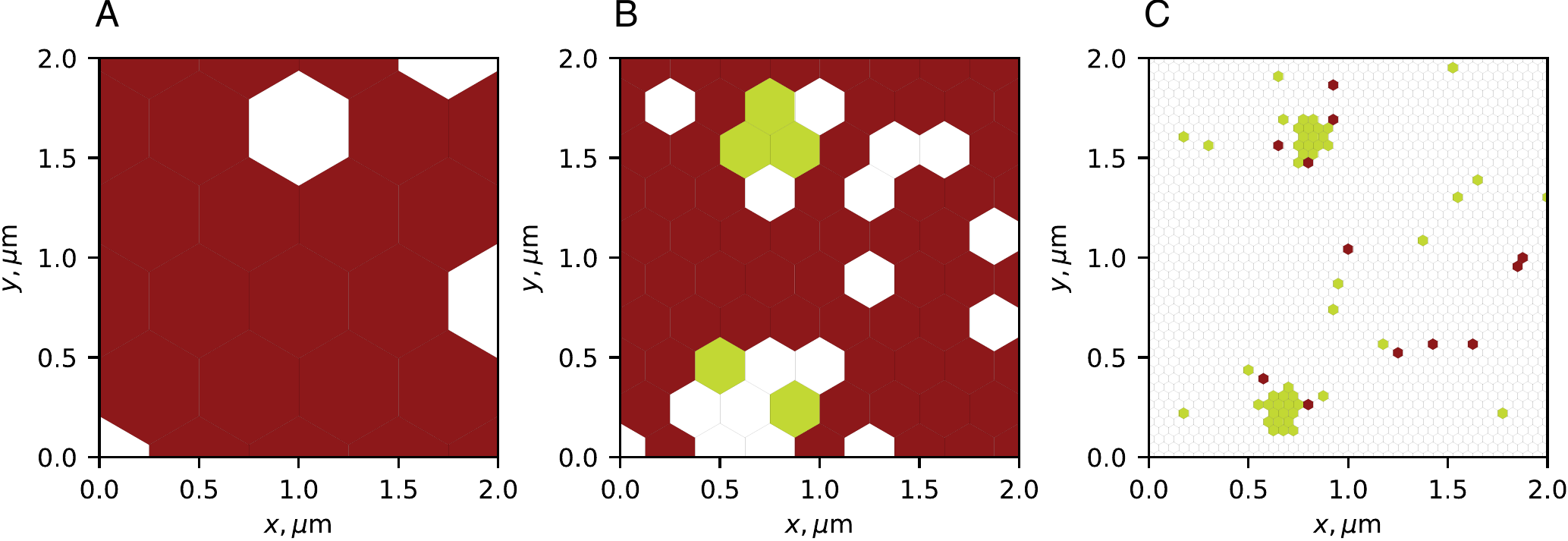}
 \caption{Thresholded Bayes factors for the VLP data set inferred at three different spatial scales.
  The average distance between bins was set to $\unit[0.5]{\mu m}$~(\textbf{A}), $\unit[0.25]{\mu m}$~(\textbf{B}), $\unit[0.05]{\mu m}$~(\textbf{C}).
  Panel~\textbf{C)} demonstrates the same mesh as in Fig.~\ref{fig:figure-tweezers-vlp}D--I.
  Color code: green --- non-spurious force, $\log_{10} K \geqslant 1$; red --- only spurious force, $\log_{10} K \leqslant -1$; white~--- insufficient evidence, $|\log_{10} K| < 1$
 }
 \label{fig:different-meshing-vlp}
\end{figure*}

As demonstrated in the simulation of Sect.~\ref{sect:beads-lattice-simulation}, the results of any inference procedure depend on how the spatial scale on which the analysis is performed, corresponds to the internal scale of the observed system.
An illustration of this fact for the VLP data can be seen in Fig.~\ref{fig:different-meshing-vlp}.
Here, our statistical test was applied to the same VLP data set on three different spatial scales.
When the bins are much larger than the typical structures present in the biological system~(Fig.~\ref{fig:different-meshing-vlp}A, $\unit[0.5]{\mu m}$), the interactions are averaged out and our statistical test confirms the absence of interactions or is inconclusive.
On the scale of the structures~(Fig.~\ref{fig:different-meshing-vlp}B, $\unit[0.25]{\mu m}$), one may identify the potential regions of interest, but is unable to resolve their internal structure.
When the bins are smaller than the regions of interest~(Fig.~\ref{fig:different-meshing-vlp}C, $\unit[0.05]{\mu m}$), given enough data, the internal structure of the regions can be resolved.
At even smaller scales~(not shown), when few points are available per bin, one starts losing the connectivity of the regions of interest, and the statistical tests becomes inconclusive or (by design) favor the model with only spurious forces~($H_0$).
We suggest that one should aim for a scale smaller than the scale of the analyzed structure, but maintain enough points per bin to reach statistically significant conclusions.

\section{Discussion}

In this paper, we introduced a method to address the inverse problem for the spatially heterogeneous OLE that is robust in regards to changes in the spurious-force contribution.
We leveraged Bayesian inference and Bayesian model comparison to account for the uncertainty in the values of the spurious force caused by a heterogeneous diffusivity field.
The method provides a test for the presence of non-diffusive forces and returns the values of the non-diffusive forces and diffusivity.

The marginalized posterior takes into account the error in the inferred forces due both to stochastic errors and to possible spurious forces when the true value of $\lambda$ is unknown. %
The expression for the Bayes factor was derived in a closed form, allowing for identification of natural parameters associated with the dynamics, namely, the signal-to-noise ratios for the total force~(drift) and spurious forces, $\ztVec$ and $\zspVec$, and the relative strength of the localization uncertainty~$4 \sigma_L^2 / (nV)$.
Interestingly, we showed that under some configurations, the discrimination between active and spurious forces is impossible without introducing additional assumptions.

As for any statistical method, a prerequisite for our method is that one observe the trajectories on the ``right'' spatial and temporal scales, which depend on the individual system.
In particular, the spatial tessellation employed here should be constructed on the appropriate spatial scale, i.e. finer than the spatial heterogeneities of interest and coarse enough to provide sufficient measurements in each mesh domain~(as discussed above).
Another condition required by our method is that the number of points per bin be~$>4$ in~1D and~$>3$ in~2D, which are the equivalent numbers of points contained in the prior.
Otherwise, due to the choice of~$\mu_\pi$, the model with only spurious forces is likely to always be favored.
Our experience with the method indicates that for the biological data we tested, $n\geqslant 20$ typically provides a reasonable compromise between the spatial and statistical resolution.

The VLP example demonstrated successful utilization of the method for the detection of biological activity.
The test was applied in an unsupervised way, which makes it useful for automatic analysis of single-molecule dynamics.
In general, however, the results may depend on the spatial meshing.
For the VLP data set, we had the advantage of \emph{a priori} knowing the characteristic spatial scale of the underlying biological processes~\autocite{Floderer2018}.
In a general case, one may need to sample multiple spatial scales in an attempt to optimize the detection.
An optimal mesh scale in this case can be seen as a trade-off between increasing statistical significance~(by getting more data per domain) and increasing resolution~(by reducing the domain size).

Potential ways to circumvent this fundamental trade-off of spatial versus statistical resolution could be to regularize the inference of the diffusivity and drift fields~\autocite{ElBeheiry2015, Laurent2019a} or to cluster the regions with similar Bayes factor values based on a certain rule.
However, the former approach induces correlations between the results inferred in different domains making analytical calculations intractable and hindering interpretation of the results.
The main difficulty with the latter approach consists in defining the appropriate clustering criterion and in accounting for how the uncertainty in the individual Bayes factors propagates to the Bayes factors of the clusters.

One should also keep in mind that the validity of the main result \fm{fm:bayes-factor-final} relies on the assumption that %
the diffusivity~$b$ is smooth enough, so that the gradient $\nabla b$ exists on the spatial scale on which the system is experimentally probed.
Additionally, we stress that $\alphaf$ and $b$ are mesoscopic quantities and their values may change depending on the analyzed scale,~\autocite[see][and the example of Sect.~\ref{sect:beads-lattice-simulation}]{Zwanzig2006}.
In practice, the choice of the spatial and temporal resolutions for the analysis is limited by the particular experimental setup and the properties of the biological system.

The Bayesian approach that we proposed here is general and not limited to the OLE equation.
The ambiguity of the stochastic integration is encountered in numerous other scientific fields involving stochastic equations with multiplicative heterogeneous noise.
The effect is usually ignored and an arbitrary standard convention is used.
The marginalized method allows us to avoid arbitrarily choosing an integral convention in the absence of system-specific information, therefore providing more robust results.

The marginalized method code is available as a module of the open source project TRamWAy~\autocite{BayesFactorsModule2018} and the microscopic crowding simulation code is available at~\autocite{Serov2019BeadsLatticeSimulation}.
Two Jupyter notebooks are provided as illustration of the module interface~\autocite{Tramway2018}.

\vskip 0.3in
\textbf{Acknowledgments.}
We thank Aleksandra Walczak, Vincent Hakim, Bassam Hajj, Mathieu Coppey and Maxime Dahan~(deceased) for helpful discussions. 
This study was funded by the Institut Pasteur, \emph{L'Agence Nationale de la Recherche}~(TRamWAy, ANR-17-CE23-0016), the INCEPTION project (PIA/ANR-16-CONV-0005, OG), and the \emph{programme d'investissement d'avenir} supported by \emph{L'Agence Nationale de la Recherche} ANR-19-P3IA-0001. 
The funding sources had no role in study design, data collection and analysis, decision to publish, or preparation of the manuscript.

\textbf{Conflicts of interest.}
The authors declare to have no financial or non-financial conflicts of interest.

\renewcommand*{\bibfont}{\footnotesize}
\setlength{\bibitemsep}{0pt}
\printbibliography

\iftoggle{NAT}{}{

\appendix

\renewcommand{\thesection}{\Alph{section}}
\renewcommand{\thesubsection}{\Alph{section}\arabic{subsection}}
\renewcommand\thefigure{{\thesection}.\arabic{figure}}
\renewcommand{\theequation}{\Alph{section}.\arabic{equation}}
\renewcommand\thetable{\thesection.\arabic{table}}
\counterwithin{figure}{section}
\counterwithin{table}{section}
\counterwithin{equation}{section}

\iftoggle{NAT}{
\maketitle
}

 \section{Appendix}

\begin{refsection}
\begingroup

\subsection{Diffusivity and diffusivity gradient inference}
\label{apx:diffusivity-inference}

The Bayes factor calculations described in the main text make use of the knowledge of the diffusivity gradient~$\g \equiv bb'$.
In theory, it is possible to incorporate the uncertainty of $bb'$ and its relation to~$b^2$ directly into the Bayes factor calculations, but it significantly complicates computations if one additionally wants to regularize~$bb'$.
The value of the spurious force is sensitive to $bb'$, so, in general, regularization is desirable.
We thus inferred the diffusivity gradient independently.
As an extra advantage, this approach keeps the method simple and allows us to obtain the results in a closed form.

The diffusivity posterior in each individual bin can be calculated by integrating the product of the likelihood~\fm{fm:likelihood} and the force-model prior~\fm{fm:H1-prior} over the non-diffusive force~$\bv a$ and~$\lambda$.
An important detail of the calculations is that we cannot use the same prior assumption~$\mu_\pi = \lambda bb' \Delta t$ that maximally favors the null model, because it requires knowing the very diffusivity gradient we are trying to calculate.
A simple solution to this problem is to assume a diffusivity-independent localization of the total force prior, for instance, $\mu_\pi = 0$.
This prior is slightly different, but it should not significantly influence the results as long as the prior is much less constraining than the likelihood.
This is the case if the number of data points in a bin~$n\gg n_\pi \sim 1$.
Under this assumption, the diffusivity posterior takes the following form:
\begin{equation*}
 p(b^2 \mid T)
 =
 \frac{
  \left(\frac{n V G'''}{2 b^2 \Delta t}\right)^{\kappa(d)}
  \exp\left(-\frac{n V G'''}{2 b^2 \Delta t}\right)
 }
 {
  b^2\, \gamma\left(\kappa(d), \frac{n V G'''}{4 \sigma_L^2}\right)
 }.
\end{equation*}
Here $G''' \equiv v + \eta^2 \zeta_t^2$, $\kappa(d) \equiv d (n+n_\pi -1)/2 -1$, $\sigma_L^2$ is the localization error and $\gamma(s, x)\equiv \int_0^x t^{s-1} \e^{-t} dt$ is the incomplete gamma function.
The posterior is normalized in 1D for $n_\pi >3$ and in 2D for $n_\pi > 2$, and the same hyper-parameter values were used as in the main text.
The maximum \emph{a posteriori} diffusivity $\mathrm{MAP}(b^2) = nV G'''/(2 \Delta t (1+\kappa))$ provided by this posterior was used to calculate the diffusivity gradient. The diffusivity gradient calculation and smoothing were performed as described in~\textcite{Laurent2019}.

\subsection{Sensitivity to hyper-parameters of the prior}
\label{apx:hyper-test}

We checked the effect of the priors' hyper-parameters $n_\pi$, $\mu_\pi$ and~$u\equiv V_\pi / V$, on the Bayes factor~\autocite{Kass1995}.
The influence of the prior was minimized by setting $n_\pi$ to the minimal value that makes the prior proper~(i.e.\ normalized).
The \emph{a priori} most likely total force~$\mu_\pi$ was chosen to maximally favor the null-model~($a=0$).
The Bayes factors dependence on the remaining parameter~$u$ is shown in Fig.~\ref{fig:bayes-factor-analytical-supp}.
The figure demonstrates that as long as the number of the observed jumps is much greater than $n_\pi$, $n \gg n_\pi$, the Bayes factors~\fm{fm:bayes-factor-final} are virtually insensitive to~$u$ both in~1D and~2D.
The interior of the magenta region~($\log_{10}\KM| \leqslant 1$) in Fig.~\ref{fig:bayes-factor-analytical-supp} corresponds to the region where a non-diffusive force will not be detected by the marginalized method.
Thus, when the number of points is comparable to the effective number of points in the prior~($n=5$), the detection sensitivity is relatively low (wide bands).
On the opposite, for a large number of points per bin~($n=100$), the sensitivity increases, and the difference between different stochastic conventions is more pronounced.
Our tests have demonstrated that in typical experimental setups, a value of $n \sim 20$ provides a reasonable balance between the detection sensitivity and mesh resolution.

\begin{figure*}[htb]
 \newcommand{\figWidth}{3.45in}
 \subfloat{
  \includegraphics[width = \figWidth]{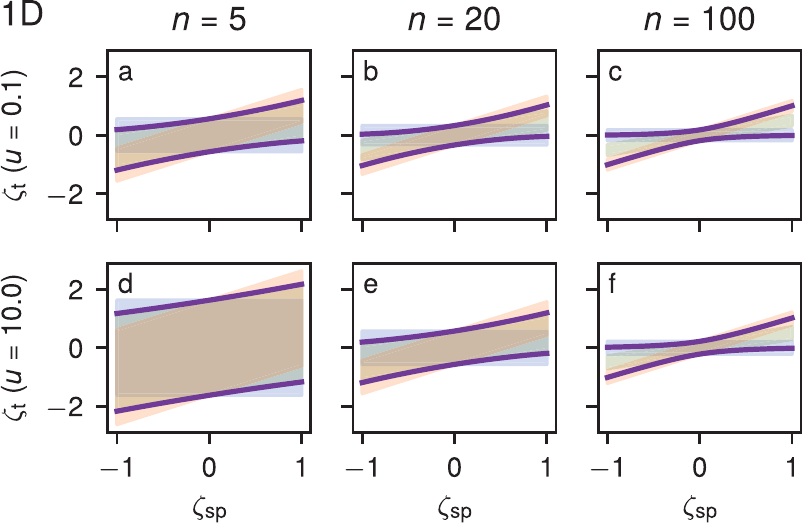}
 }
 \subfloat{
  \includegraphics[width=\figWidth]{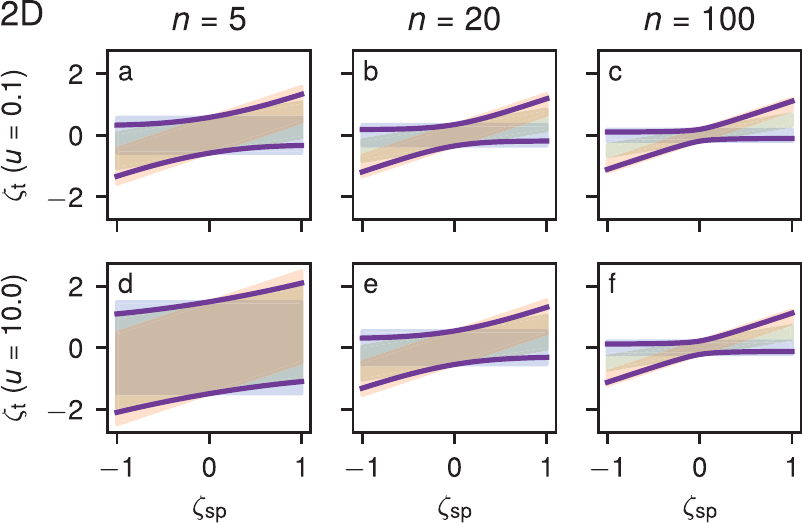}
 }
 \caption{%
  Bayes factor plots for different~$u\equiv V_\pi / V$ and~$n$ in~1D and~2D.
  Shaded regions show the $|\log_{10}\KF| \leqslant 1$ region for fixed-$\lambda$ conventions: It\^o~(blue), Stratonovich~(green), H\"anggi~(orange).
  The solid magenta lines are the $\log_{10} \KM = 1$ isopleths for the marginalized method.
  Parameters: $n_\pi = 4$ for~1D, $n_\pi = 3$ for~2D
 }
 \label{fig:bayes-factor-analytical-supp}
\end{figure*}

\subsection{Undetectable forces}
\label{apx:undetectable-forces}

Here, we describe an interesting property showing that statistical testing may fail to detect forces even when an arbitrarily large number of data points is available.

In 1D, any force satisfying $0 \leqslant \zt/\zsp \leqslant 1$ will not be detected by the marginalized method for any $n$.
To demonstrate this, we perform the substitutions $R \equiv \zt / \zsp$, $q\equiv R - \lambda$ and~$q'\equiv \eta(R - \lambda)$ in the Bayes factor~\fm{fm:bayes-factor-final} to obtain
\begin{equation*}
 \KM
 \equiv
 \frac
 {
  \int_{\eta(R-1)}^{\eta R} dq'\ (v + \zsp^2 q'^2)^{-\power(d)}
 }
 {
  \int_{R-1}^{R} dq\ (v + \zsp^2 q^2)^{-\power(d)}
 }.
\end{equation*}
Both integrals of Eq.~\fm{fm:bayes-factor-final} now have the same form, but different integration intervals.
Since $0 < \eta < 1$ and $0 \leqslant R \leqslant 1$, it then follows that $R-1 \leqslant \eta (R-1) \leqslant 0 \leqslant \eta R \leqslant R$ and thus that
the interval $[\eta (R - 1); \eta R]$ is always contained in the interval $[R - 1; R]$.
Given that $(v + \zsp^2 q^2) ^{-\power(d)} > 0$, this leads to $\KM \leqslant 1$.
Hence, the force model~$H_1$ will never be favored by the Bayes factor for $0 \leqslant R \leqslant 1$, no matter how large $n$ is.

In 2D, the presence of an orthogonal force component qualitatively changes the behavior of~$\KM$.
If $\ztper = 0$, the 1D results stands.
If~$\ztper \neq 0$, directly taking the limit of Eq.~\fm{fm:bayes-factor-final} yields $\lim_{n\to\infty} \KM = +\infty$.
Hence in the presence of an orthogonal total force component, a force can always be identified given a large enough number of observations.
The asymptotic undetectable region in 2D is thus a 1D interval $0 \leqslant \ztpar/\zsp \leqslant 1$, $\ztper = 0$.

In practice, the closer the value of $\ztper$ is to~0, the more data points $n$ are necessary to discriminate between~$H_0$ and~$H_1$.
Conversely, collecting more points leads to a higher measurement accuracy for $\ztper$, which may bring it even closer to~0.
Hence, in practice, the undetectable region in 2D will rather be a 2D region around~$\{0 \leqslant \ztpar/\zsp \leqslant 1; \ztper = 0\}$, the width of which will be defined by the available number of data points.

\subsection{Localization uncertainty and motion blur}
\label{apx:positioning-noise}

We show here how to include the effects of positional noise and motion blur into the inference scheme and the statistical tests.
The main sources of localization uncertainty are photon shot noise due to diffraction in the microscope, and motion blur due to finite camera shutter time~\autocite{Vestergaard2016ExperimentalParameters}.
The motion blur also affects the covariance of the recorded displacements and~--- together with the localization error~--- leads to the following apparent particle diffusivity:~$\tilde{b}^2 = 2/3 b^2 + 2\sigma_L^2/ \Delta t$ when the camera shutter is kept continuously open~\autocite{Savin2005,Berglund2010,Vestergaard2016}.
Here $b^2$ is the true particle diffusivity and $\sigma_L^2$ is the variance of the localization error. %
The apparent bias can be corrected for by using the method-of-moments scheme~\autocite{Vestergaard2014}.
For the calculation of the Bayes factor, one must also take into account that the domain for $\tilde{b}^2$ is $[2\sigma_L^2 / \Delta t; \infty)$ rather than $[0; \infty)$ as for $b^2$.
This is done by a renormalization of the priors $\pi(\bv a, \lambda, \bv g, b^2)$ and $\pi(\lambda, \bv g, b^2)$ leading to
\begin{strip}
\begin{equation}
 \begin{aligned}
  \KM
  &=
  \eta^d
  \frac{
   \int_0^1 d \lambda \;
   \gamma\left(
   \power(d), \frac{n V}{4 \sigma_L^2} G(\lambda)
   \right)
   G^{-\power(d)}(\lambda)
  }
  {
   \int_0^1 d \lambda \;
   \gamma\left(
   \power(d), \frac{n V}{4 \sigma_L^2} H(\lambda)
   \right)
   H^{-\power(d)}(\lambda)
  },
&\quad
  \KF
  &=
  \eta^d
  \frac{
   \gamma\left(
   \power(d), \frac{n V}{4 \sigma_L^2} G(\lambda)
   \right)
  }
  {
   \gamma\left(
   \power(d), \frac{n V}{4 \sigma_L^2} H(\lambda)
   \right)
  }
  \left[
   \frac{
    G(\lambda)
   }
   {
    H(\lambda)
   }
   \right]^{-\power(d)},
 \end{aligned}
 \label{fm:bayes-factor-with-localization-uncertainty}
\end{equation}
\end{strip}
where $\gamma(s, x)\equiv \int_0^x t^{s-1} \e^{-t} dt$ is, as before, the incomplete gamma function, $G(\lambda) \equiv v + \eta^2(\ztVec - \lambda \zspVec)^2$, and $H(\lambda) \equiv v + (\ztVec - \lambda \zspVec)^2$.
The sample variance $V$ already includes a contribution from the localization error and does not need to be updated.
Note also the new dimensionless parameter $4 \sigma_L^2 / (nV)$ that characterizes the relative strength of the localization uncertainty.
Increasing the number of observations $n$ will reduce the effect of the localization error.

Including the localization error and motion blur in the force posterior~\fm{fm:active-force-posterior} gives
\begin{strip}
\begin{equation}
 p(\bv \zeta_a \mid T)
 =
 (\pi(1-\eta^2))^{-d/2}
 \frac{
  \int_0^1 d\lambda \
  G'(\lambda)^{-\kappa(d) - d/2}\
  \gamma\left(\kappa(d) + d/2, \frac{nV}{4 \sigma_L^2} G'(\lambda)\right)
 }
 {
  \int_0^1 d\lambda \
  G(\lambda)^{-\kappa(d)}\
  \gamma\left(\kappa(d), \frac{nV}{4 \sigma_L^2}G(\lambda)\right)
 },
 \label{fm:active-force-posterior-with-error}
\end{equation}
\end{strip}
where
$G'(\lambda) \equiv
 v + \eta^2/(1-\eta^2)\zeta_a^2 + (\ztVec - \zaVec -  \lambda \zspVec)^2$.
Note that this is a joint distribution for all force components.
We have also included here the normalization coefficient which was not shown in the main text.
The posteriors for fixed-$\lambda$ conventions are obtained by dropping the $\lambda$~integrals.

The posterior for a single component~($x$ or~$y$) of the force in 2D is obtained by integrating Eq.~\fm{fm:active-force-posterior-with-error} over the other component.
For example, for the $x$-component of the force, one gets
\begin{strip}
\begin{equation*}
 p(\zeta_{ax} \mid T)
 =
 \sqrt
 \frac{n+n_\pi}n
 \frac{
  \int_0^1 d\lambda \
  G''(\lambda)^{-\kappa(d) - d/2}\
  \gamma\left(\kappa(d) + d/2, \frac{nV}{4 \sigma_L^2} G''(\lambda)\right)
 }
 {
  \int_0^1 d\lambda \
  G(\lambda)^{-\kappa(d)}\
  \gamma\left(\kappa(d), \frac{nV}{4 \sigma_L^2}G(\lambda)\right)
 },
 \label{fm:active-force-posterior-with-error-x}
\end{equation*}
\end{strip}
where $d=2$ and
$G''(\lambda) \equiv
 v
 + n_\pi/n \zeta_{ax}
 + (\zeta_{tx} - \zeta_{ax} - \lambda \zeta_{spx})^2
 + n_\pi/(n+n_\pi) (\zeta_{ty} - \lambda \zeta_{spy})^2$.
This expression was used for the posterior plots in Fig.~\ref{fig:figure-tweezers-vlp}B of the main text and in Appendix~\ref{apx:non-symmetric-posterior}.

\subsection{Non-symmetric marginalized posterior}
\label{apx:non-symmetric-posterior}

The force posterior of the marginalized method does not have the reflection symmetry that fixed-$\lambda$ posteriors do~(Fig.~\ref{fig:active-force-posteriors-apx}).
This property stems directly from our choice to center the prior around~$\zaVec = 0$ (no non-diffusive force).
This results in a symmetric total force prior centered around~$\ztVec = \lambda \zspVec$, which, in general, yields a non-symmetric posterior once integrated over~$\lambda$.
More generally, any $\lambda$-independent force prior centered around $\zaVec = \mathrm{const}(\lambda)$ will result in an asymmetric posterior, and the asymmetry can be significant even when~$\sqrt{n_\pi/n} \ll 1$.
The asymmetry is observed both in 1D and~2D.

\begin{figure}[htb]
 \includegraphics[width = 3.5in]{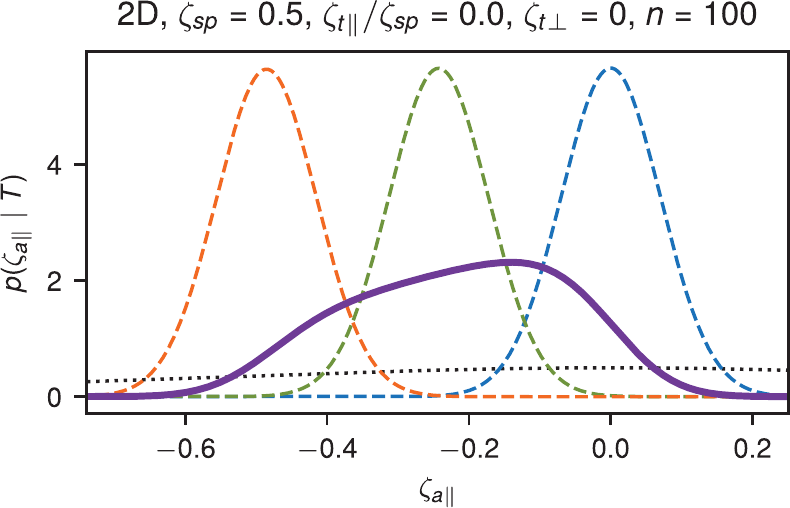}
 \newcommand{\aPar}{a_\parallel}
 \caption{%
  Force posteriors for a system, in which the asymmetric character of the marginalized posterior is evident.
  The plot shows the posterior distribution of the signal-to-noise ratio~$\zapar \equiv \aPar \Delta t / \sqrt{V}$ for the non-diffusive force component $\aPar$ parallel to the spurious force~$\zspVec$.
  The parameters used: $\zsp = 0.1$, $\ztVec = 0$, $n = 500$ and no localization error~($\sigma_L^2=\unit[0]{\mu m^2}$).
  Probability distributions obtained for It\^o (dashed blue), Stratonovich (dashed green), H\"anggi (dashed orange) and marginalized (solid magenta) approaches are shown alongside their common prior (black dotted).
  For the prior plot, we assumed that $V_\pi = V$, so that~$\zapar = \aPar \Delta t / \sqrt{V_\pi}$
 }
 \label{fig:active-force-posteriors-apx}
\end{figure}

\subsection{Numerical Simulations}
\label{apx:numerical-simulation-parameters}
Numerical trajectories were generated in a \unit[1]{$\mu$m} $\times$ \unit[1]{$\mu$m} box with periodic boundary conditions.
We created a piecewise-constant diffusivity gradient $g=\pm \unit[0.02]{\mu m/s}$ by setting a symmetric triangular diffusivity profile along the $x$~axis: $D(x, y) = D_{max} - 2(D_{max} - D_{min}) |x - 0.5|$, with $D_{min} = \unit[0.01]{\mu m^2/s}$, $D_{max} = \unit[0.02]{\mu m^2/s}$~(Fig.~\ref{fig:figure-2}B).
This setup provides a continuous diffusivity, which avoids creating boundary artifacts.
Each trajectory consisted of $N=10^4$ jumps with a time step between the recordings of~$\Delta t = \unit[0.04]{s}$.
To make sure the discrete nature of the simulations does not influence the analysis, the internal simulation time step was set to $\Delta t/100$, and the final data set was sampled from this data set with $\Delta t$.
For the given simulation parameters, setting $\Delta t/100$ was sufficient, as can be seen in Fig.~\ref{fig:internal-time-step-comparison}, where further decrease to $\Delta t/1000$ can be seen to yield the same results.

\begin{figure*}[htb]
 \includegraphics[width = 7in]{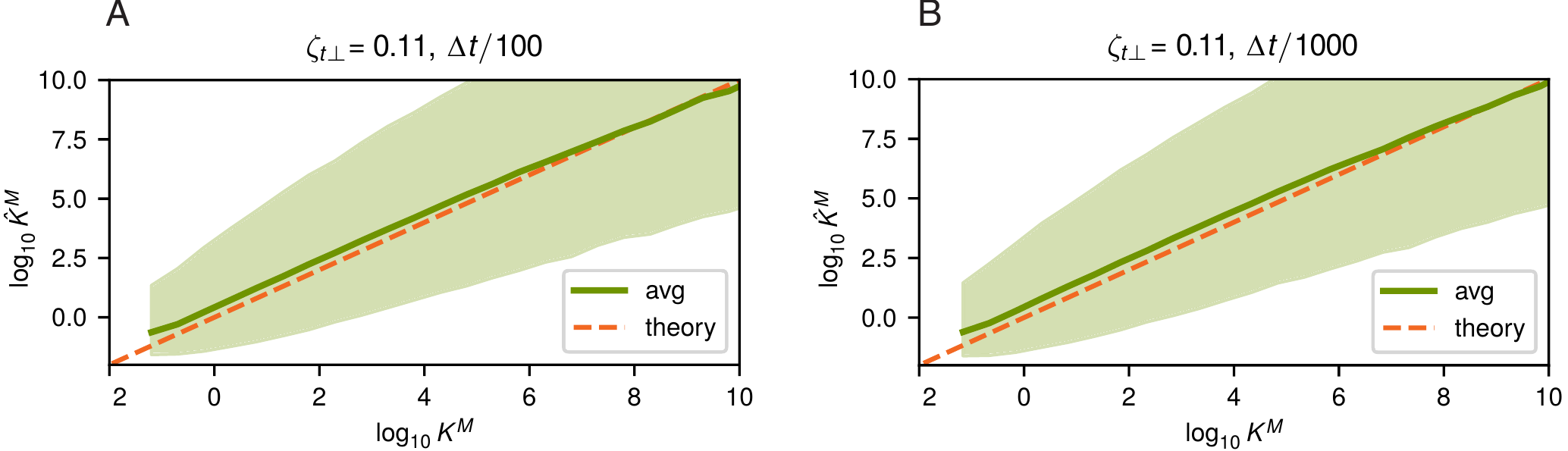}
 \caption{%
  Statistical performance of the proposed marginalized method calculated with 100~(\textbf{A}) and 1000~(\textbf{B}) internal time steps.
  Both panels were simulated with a perpendicular total force component~$\ztper = 0.11$.
  Panel~\textbf{A} shows the same plot as in Fig.~\ref{fig:figure-2}D
 }
 \label{fig:internal-time-step-comparison}
\end{figure*}

A uniform total force was imposed throughout the system.
Simulations were performed 100 times for each different value of the parallel~($x$) component of the total force in the range $-25 \leqslant \ztpar / \zsp \leqslant 25$, and for two different values of its $y$-component: $\ztper = 0$ (parallel spurious and total forces) and $\ztper = 0.11$.

Tessellation of the region was performed using an adaptive ``grow when required'' neural gas algorithm~\autocite{Marsland2002}, with a minimum count of 20~jumps per domain~(bin).
The average distance between the initial cell centers was set to 2 times the mean absolute jump length across the whole region.
For each domain, the prior variance $V_\pi$ was set equal to the variance of the recorded displacements averaged over all but the current domain.

\subsection{Microscopic crowding simulations}
\label{apx:beads-lattice-simulations}
This section describes the parameters of the numerical simulations of diffusion in-between a lattice of immobile beads.
The goal of the simulation was to illustrate how the diffusivity gradient and spurious forces observed on the mesoscopic scale in biological systems can be due to crowding on the microscopic scale.
In this perspective, the simulations were performed with nano-meter-scale obstacles, while the analysis was performed on the micron-scale.

The simulated 2D region was a $\unit[10 \times 10]{\mu m^ 2}$ box with periodic boundary conditions along both axes.
The region contained immobile impenetrable beads located in the nodes of a regular square lattice at a distance of $\Delta x = \Delta y = \unit[0.04]{\mu m}$ between the nodes.
The radius of the beads changed linearly along the $x$-axis from $r = \unit[1]{nm}$ to $r = \unit[15]{nm}$ following a triangular profile according to the following formula: $r(x) = 15 - 2.8 |x-5|$.
$r(x)$ did not depend on the $y$ coordinate.
The triangular profile was used to avoid discontinuities in the diffusivity.
The size of the obstacles was chosen to resemble the scale of small individual obstacles in the cell cytoplasm.
In the cytoplasm of a real cell, the other cytoplasm constituents are not immobile and also participate in the thermal and directed movement, but for the sake of the illustration the system was simplified.

To probe this environment, a trajectory of a mobile particle experiencing free (no-force) diffusion with diffusivity $D=\unit[1]{\mu m^2}$ within this environment was simulated.
On the surface of the beads, the particle experienced a perfectly elastic reflection.
Each simulated trajectory contained 1000~jumps with a time step of $\Delta t = \unit[0.4]{ms}$.
In total, to sample the whole region, 1000~independent trajectories were generated, with the origin points uniformly distributed over the area unoccupied by the immobile beads.
Due to the absence of non-diffusive forces, the system is in physical equilibrium, so in theory, the correct value of~$\lambda$ is known~($\lambda = 1$).

To perform the inference procedure, all points of the trajectories were grouped in bins of a hexagonal mesh with a distance of of $R_{bin} \approx \unit[0.5]{\mu m}$ between bin centers.
Thus, each individual bin included around $\pi (R_{bin}/2)^2 / (\Delta x \Delta y) \approx 120$ immobile beads.
As a consequence, the effects of the excluded space and reflections on the microscopic scale were perceived as spatial diffusivity variations on the mesoscopic scale of~$R_{bin}$.
Meanwhile, the size of the whole system was chosen large enough to feature diffusivity changes on the scale of many bins along each axis: $L/(R_{bin}) = 20$.
The full simulation and analysis code is available on GitHub~\autocite{Serov2019BeadsLatticeSimulation}.
The localization uncertainty was set to~$\sigma=\unit[0]{\mu m}$.

\endgroup

\normalem
\renewcommand*{\bibfont}{\footnotesize}
\setlength{\bibitemsep}{0pt}
\printbibliography[heading=subbibliography]
\end{refsection}
}

\end{document}